\documentclass[twocolumn,tighten]{aastex62}
\turnoffedit 
\accepted{on August 10th, 2018}
\setwatermarkfontsize{0.5in}

\usepackage[us,12hr]{datetime}
\usepackage{graphicx}
\usepackage{gensymb}
\usepackage{listings}
\usepackage{color}
\usepackage{mathrsfs}
\usepackage{natbib}
\usepackage{booktabs}
\usepackage{comment}
\usepackage{morefloats}
\usepackage{sidecap}
\usepackage{grffile}
\usepackage{verbatim}
\usepackage{longtable}
\usepackage{etoolbox}
\usepackage{amsmath}
\usepackage{amssymb}
\usepackage{catchfilebetweentags}
\usepackage{multirow}

\newcommand{\kms}{\mbox{$\rm km\,s^{-1}$}}
\newcommand{\Kkmsec}{\mbox{$\rm K\,km\,s^{-1}$}}

\newcommand{\msun}{\mbox{$\rm M_{\odot}$ }}

\newcommand{\msunyr}{\mbox{$M_{\odot}\,{\rm yr}^{-1}$}}

\newcommand{\lsun}{\mbox{$L_{\odot}$}}
\newcommand{\unitaco}{\mbox{${M_{\odot}}\,(\Kkmsec\,{\rm pc}^{2}$)$^{-1}$}}

\newcommand{\lir}{\mbox{$L_{\rm IR}$}}

\newcommand{\aco}{\mbox{$\alpha_{\rm CO}$}}

\newcommand{\cohh}{CO$\to$H$_2$}

\newcommand{\coone}{\mbox{${\rm CO\,{\it J}=1\to0}$}}

\newcommand{\coseven}{\mbox{${\rm CO\,{\it J}=7\to6}$}}

\newcommand{\citwo}{\mbox{{\rm [\ion{C}{1}]$\,{{^3P_2}\to{^3P_1}}$}}}
\newcommand{\water}{\mbox{{\rm H$_{2}$O\,${2_{11}}\to{2_{02}}$}}}

\newcommand{\herschel}{{\it Herschel}}

\newcommand{\hone}{\mbox{\rm \ion{H}{1}}}
\newcommand{\htwo}{\mbox{\rm H$_2$}}

\newcommand{\ci}{\mbox{\rm \ion{C}{1}}}

\newcommand{\clean}{{\scshape clean}}

\makeatletter
\def\CatchFBT@Fin@l#1[#2]{   \begingroup
            \makeatletter #2      \scantokens\expandafter{         \expandafter\CatchFBT@tok\expandafter{\the\CatchFBT@tok}}      \CatchFBT@IsAToken{#1}
         {\global#1\expandafter{\the\CatchFBT@tok}}
         {\xdef#1{\the\CatchFBT@tok}}      \ifx\CatchFBT@tok#1\else\global\CatchFBT@tok{}\fi
   \endgroup
}\makeatother

\graphicspath{{./figures/}}
\makeatletter
\def\input@path{{./tables/}}
\makeatother

\newcommand{\tr}{{\tt TiRiFiC}}
\newcommand{\radex}{{\tt RADEX}}
\newcommand{\emcee}{{\tt emcee}}
\newcommand{\objfull}{HerMES\,J022016.5$-$060143}
\newcommand{\obj}{HXMM01}

\defcitealias{Fu:2013hz}{Paper\,I}
\shorttitle{HXMM01}
\shortauthors{Xue et al.}

\begin{document}

\title{Flat Rotation Curves found in Merging Dusty Starbursts at $z=2.3$ through Tilted-Ring Modeling}

\author[0000-0001-7689-9305]{Rui Xue}
\affiliation{Department of Physics \& Astronomy, University of Iowa, 203 Van Allen Hall, Iowa City, IA 52242, USA}

\author[0000-0001-9608-6395]{Hai Fu}
\affiliation{Department of Physics \& Astronomy, University of Iowa, 203 Van Allen Hall, Iowa City, IA 52242, USA}

\author[0000-0002-1272-6322]{Jacob Isbell}
\affiliation{Department of Physics \& Astronomy, University of Iowa, 203 Van Allen Hall, Iowa City, IA 52242, USA}

\author[0000-0001-5118-1313]{R. J.\ Ivison}
\affiliation{European Southern Observatory,
  Karl-Schwarzchild-Stra{\ss}e 2, D-85748 Garching, Germany}
\affiliation{Institute for Astronomy, University of Edinburgh,
  Royal Observatory, Blackford Hill, Edinburgh EH9 3HJ, UK}
  
\author[0000-0002-3892-0190]{Asantha Cooray}
\affiliation{Department of Physics \& Astronomy, University of California, Irvine, CA 92697, USA}

\author[0000-0001-5875-3388]{Iv\'an Oteo}
\affiliation{Institute for Astronomy, University of Edinburgh,
  Royal Observatory, Blackford Hill, Edinburgh EH9 3HJ, UK}
\affiliation{European Southern Observatory,
  Karl-Schwarzchild-Stra{\ss}e 2, D-85748 Garching, Germany}

\begin{abstract}
The brightest 500$\,\mu$m source in the XMM field, HXMM01, is a rare merger of luminous starburst galaxies at $z=2.3$ with a dust-obscured star-formation rate of 2,000$\,M_{\odot}\,{\rm yr}^{-1}$. Here we present high-resolution spectroscopic observations of HXMM01 with the Atacama Large Millimeter/submillimeter Array (ALMA). We detect line emission from ${\rm CO\,{\it J}=7\to6}$, [C I]$\,{{^3P_2}\to{^3P_1}}$, and p-${\rm H_{2}O}\,{2_{11}}\to{2_{02}}$ and continuum emission at $230\,$GHz. At a spatial resolution of 0\farcs2 and a spectral resolution of 40$\,\rm km\,s^{-1}$, the source is resolved into three distinct components, which are spatially and dynamically associated within a projected radius of 20$\,$kpc and a radial velocity range of 2,000$\,\rm km\,s^{-1}$.
For two major components, our Bayesian-based tilted-ring modeling of the ALMA spectral cubes shows almost flat rotation curves peaking at $\sim500\,\rm km\,s^{-1}$ at galactocentric distances between 2 and 5$\,$kpc. Each of them has a dynamical mass of $\sim10^{11}\,M_\odot$.
The combination of the dynamical masses and the archival ${\rm CO\,{\it J}=1\to0}$ data places strong upper limits on the CO$\to$H$_2$ conversion factor of $\alpha_{\rm CO}\lesssim1.4-2.0\,{M_{\odot}}\,\rm (K\,km\,s^{-1}\,pc^{2})^{-1}$. 
These limits are significantly below the Galactic inner disk $\alpha_{\rm CO}$ value of $4.3\,{M_{\odot}}\,\rm (K\,km\,s^{-1}\,pc^{2})^{-1}$ but are consistent with those of local starbursts. 
Therefore, the previously estimated short gas depletion timescale of $\sim200\,$Myr remains unchanged.
\end{abstract}

\section{Introduction}\label{sec:intro}

The launch of the {\it Herschel}\footnote{\herschel\ is an ESA space observatory with science instruments provided by European-led Principal Investigator consortia and with important participation from NASA.} Space Observatory \citep{Pilbratt:2010en} allowed us to identify a rare population of extremely infrared-bright ($S_{\rm500\mu m}>100$\,mJy) sources at redshifts of $z\approx1-3$.
Although the population is dominated by gravitationally lensed galaxies \citep[e.g.,][]{Negrello:2010fx,Fu:2012iu,Wardlow:2013ie,Calanog:2014ep,Bussmann:2015jx,Nayyeri:2016iy,Negrello:2017jp},
a small fraction of these sources ($<10\%$) are expected to be intrinsically hyperluminous \citep[$L_{\rm IR}\geq10^{13}\,L_\odot$, e.g.,][]{Fu:2013hz,Ivison:2013hj}.
Similar to the submillimeter-bright galaxies \citep[SMGs;][]{Smail:1997kh,Barger:1998em,Hughes:1998gn}, these hyperlumious IR galaxies (HyLIRGs) are likely caught in a short-lived starburst phase. 
The molecular gas reservoir of the disks cannot sustain the extreme star formation rate for more than $\sim$200\,Myr \citep[e.g.,][]{Bothwell:2013ed}. 

\objfull\ (a.k.a., \obj) is a spectacular example of this hyperluminous population \citep[][hereafter \citetalias{Fu:2013hz}]{Fu:2013hz}. In our previous observations, the bright \herschel\ source ($S_{\rm500\mu m}=132$\,mJy) at $z = 2.308$ is resolved into a merging pair of gas-rich starburst galaxies separated by 
3\arcsec\ (or a projected distance of 25\,kpc).
Both components are only mildly magnified ($\mu\approx1.6$) by a pair of foreground galaxies.
The intrinsic IR luminosity of $2\times10^{13}$\,\lsun\ makes it one of the most luminous unlensed SMGs. Despite the broad H$\alpha$ lines, the panchromatic SEDs show no evidence of active galactic nuclei (AGN), in contrast to other AGN-dominiated HyLIRGs \citep[e.g.,][]{Ivison:1998iw,Ivison:2010kz}. Our 77\,ks Chandra ACIS-S observations (Obs-ID: 14972) did not detect any significant X-ray emission at the location of \obj. 
The upper limits of $0.5-8$\,keV X-ray luminosity ($2.80\times10^{43}$ and $3.86\times10^{43}$ erg\,s$^{-1}$ for the northern and southern component, respectively) are consistent with the expectations from X-ray binaries, based on the $L_{\rm X}$$-$SFR relation of \citet{Mineo:2012kz}. 

Despite the extensive data set presented in \citetalias{Fu:2013hz}, the resolved kinematic structures of \obj\ remain to be determined to understand the physical mechanism(s) driving the prolific star formation.
In particular, spatially revolved kinematics is a powerful tool to determine the mass distribution of baryonic and dark matter \citep[e.g.,][]{Noordermeer:2007gm,deBlok:2008cn,Swaters:2009ku,Lang:2017bv}, and can also constrain the much debated \cohh\ conversion factor\footnote{The mass from Helium is included in the definition here, with $M_{\rm mol}=1.36M_{\rm H_2}$.} ($\alpha_{\rm CO} \equiv M_{\rm mol}/L_{\rm CO}$) in high-redshift galaxies \citep[e.g.,][]{Ivison:2011it,Genzel:2012fw,Hodge:2012jt,Magnelli:2012br,Narayanan:2012fa}.
In this {\it Letter}, we present 0\farcs2-resolution gas kinematics of \obj\ traced by two molecular lines and one atomic line from observations with the Atacama Large Millimeter/submillimeter Array (ALMA). In \S\,\ref{sec:obs}, we describe the observations and our data processing procedures. In \S\,\ref{sec:results}, we present the observational results, the kinematic models, and the derived rotation curves, dynamical masses, and constraints on $\alpha_{\rm CO}$. In \S\,\ref{sec:discussion}, we discuss the implications of our findings. Throughout we adopt the concordance $\Lambda$CDM cosmology with $\Omega_{\rm m}=0.27$, $\Omega_\Lambda=0.73$, and $H_0$ = 70\,\kms\,Mpc$^{-1}$ \citep{Komatsu:2011in}. 

\section{Observations and Data}\label{sec:obs}

\subsection{ALMA Band-6 Observations}\label{sec:alma}

\floattable
\begin{deluxetable}{cccrc}[htb!]
\tabletypesize{\small}
\tablecolumns{7} 
\tablecaption{Properties of Observed Lines\label{tb:lines}}
\tablewidth{0pt} 
\tablehead{
\multicolumn{1}{c}{Species}&
\multicolumn{1}{c}{Transition}&
\multicolumn{1}{c}{Rest-Freq.}&
\multicolumn{1}{c}{$E_{\rm up}/k$}&
\multicolumn{1}{c}{$n_{\rm crit}$}\\
\multicolumn{1}{c}{}&
\multicolumn{1}{c}{$$}&
\multicolumn{1}{c}{GHz}&
\multicolumn{1}{c}{K}&
\multicolumn{1}{c}{cm$^{-3}$}
} 
\startdata
\toprule
$p$-H$_{2}$O& ${2_{11}}\to{2_{02}}$ & 752.03314 & 136.9 & $2.1\times10^{7}$ \\
CO		    & $J=7\to6$             & 806.65181 & 154.9 & $1.2\times10^{5}$ \\
\ion{C}{1}  & ${^3P_2}\to{^3P_1}$   & 809.34197 & 62.5 	& $1.3\times10^{3}$ \\
\midrule
CO		    & $J=1\to0$             & 115.27120 & 5.5 	& $3.2\times10^{2}$ \\
\bottomrule
\enddata
\tablecomments{The critical densities are calculated as $n_{\rm crit}=\frac{\Sigma_{i>j} A_{ij}}{\Sigma_{i\neq j}\gamma_{ij}}$, using the coefficients from the Leiden Atomic and Molecular Database \citep[LAMDA,][]{Schoier:2005wx}. We set the kinetic temperature to $T_{\rm kin}=50$\,K and consider only \htwo\ molecules (an ortho-to-para abundance ratio of 3) as the collisional partner.
}
\end{deluxetable}

ALMA band-6 observations of \obj\ were carried out on 2016 August 2, 14, 15, and 17 under the cycle-3 project 2015.1.00723.S. We tuned three 2\,GHz spectral windows to the redshifted frequencies of the \coseven, \citwo, and \water\ lines between 226 and 245\,GHz (see Table\,\ref{tb:lines}), and used an additional 2\,GHz window to cover a line-free continuum region centered at $\nu_{\rm obs}=230$\,GHz. Each window had an effective bandwidth of 1875\,MHz and a channel spacing of 15.625\,MHz. The total on-source integration time was 2.6\,hr, with thirty-eight to forty-five antennas online in the C40-5 configuration. The observations consisted of a single pointing towards the approximate center of \obj\ ($\alpha_{\rm J2000}$=$02^{\rm h}20^{\rm m}16\farcs613$, $\delta_{\rm J2000}$=$-06\arcdeg01\arcmin43\farcs15$). The variations in amplitude and phase were calibrated using J0241$-$0815. The bandpass and flux density calibrators are J0238$+$1636 and J0006$-$0623, respectively.

The raw data were calibrated using the ALMA pipeline in the Common Astronomy Software Application \citep[CASA;][]{McMullin:2007ta}. We performed the $uv$-plane continuum subtraction and data imaging in CASA\,ver.\,5.1.2, using the {\tt mstransform} and {\tt tclean} tasks, respectively. We used the Briggs image weighting scheme with {\tt robust = 0} to suppress sidelobes. The synthesized beam at the \coseven\ frequency ($\nu_{\rm obs} = 243.8$\,GHz) is 0\farcs24$\times$0\farcs18 with P.A.=84\arcdeg; so we set a pixel size of 0\farcs03 to oversample the beam. For image deconvolution, we adopted the multi-scale {\sc clean} algorithm implemented in {\tt tclean}, and applied a circular \clean\ mask with a radius of 10\arcsec\ centered at \obj, where its emission is expected. 
Due to the default Hanning weighting function applied online, the spectral resolution is twice the channel spacing and the noise is correlated between adjacent visibility channels \footnote{The effective noise bandwidth of each channel is $2.667\times$ channel spacing (\url{https://safe.nrao.edu/wiki/bin/view/Main/ALMAWindowFunctions})}. 
We thus set a channel width of 40\,\kms\ for spectral line imaging, which is equivalent to the resolution FWHM.
Our imaging products consist of two maps per line/continuum:
a ``data'' map which is corrected for the primary beam response of ALMA 12\,m antenna, and an ``uncertainty'' map providing the estimated $1\sigma$ noise.
The $1\sigma$ noise at the center of the 230\,GHz continuum map reaches $\sim0.02$\,mJy\,beam$^{-1}$, consistent with expectation.

\subsection{Archival \coone\ Data}

The \coone\ data of \obj\ were obtained with the Karl G. Jansky Very Large Array (VLA) in the DnC, C, and B configurations in 2012 (Program IDs: 11B-044 and 12A-201).
The phase and amplitude variations were calibrated by observing J0241-0815, and 3C48 was used as the bandpass and flux density calibrator.
The total on-source integration time is 3.8\,hr. \citetalias{Fu:2013hz} presented the earlier data product of the same observations, in which the reduction was performed in {\tt AIPS} with a slightly different flux scaling for 3C\,48: $S_{\rm 34.8\,GHz}=0.83$\,Jy; i.e., 5\% higher than the value adopted in CASA ($S_{\rm34.8\,GHz}=0.79$\,Jy; \citealt{Perley:2013fa}). We reprocessed the VLA \coone\ data in CASA\,ver.\,5.1.2 to allow a better comparison with the ALMA band-6 wdata. We performed imaging with {\tt tclean} as in \S\,\ref{sec:alma}. The synthesized beam is 0\farcs54$\times$0\farcs51 with P.A.$=-67.91\arcdeg$ ({\tt robust = 0.5}). The \coone\ datacube is sampled with a pixel size of 0\farcs06 and a channel width of 75\,\kms.

\section{Results}
\label{sec:results}

\subsection{Detection of {\rm CO}, {\rm [\ci]}, and {\rm H$_{2}$O}}
\label{sec:linedetect}
\label{sec:results:line}

\begin{figure*}\centering
\includegraphics[width=1.0\textwidth]{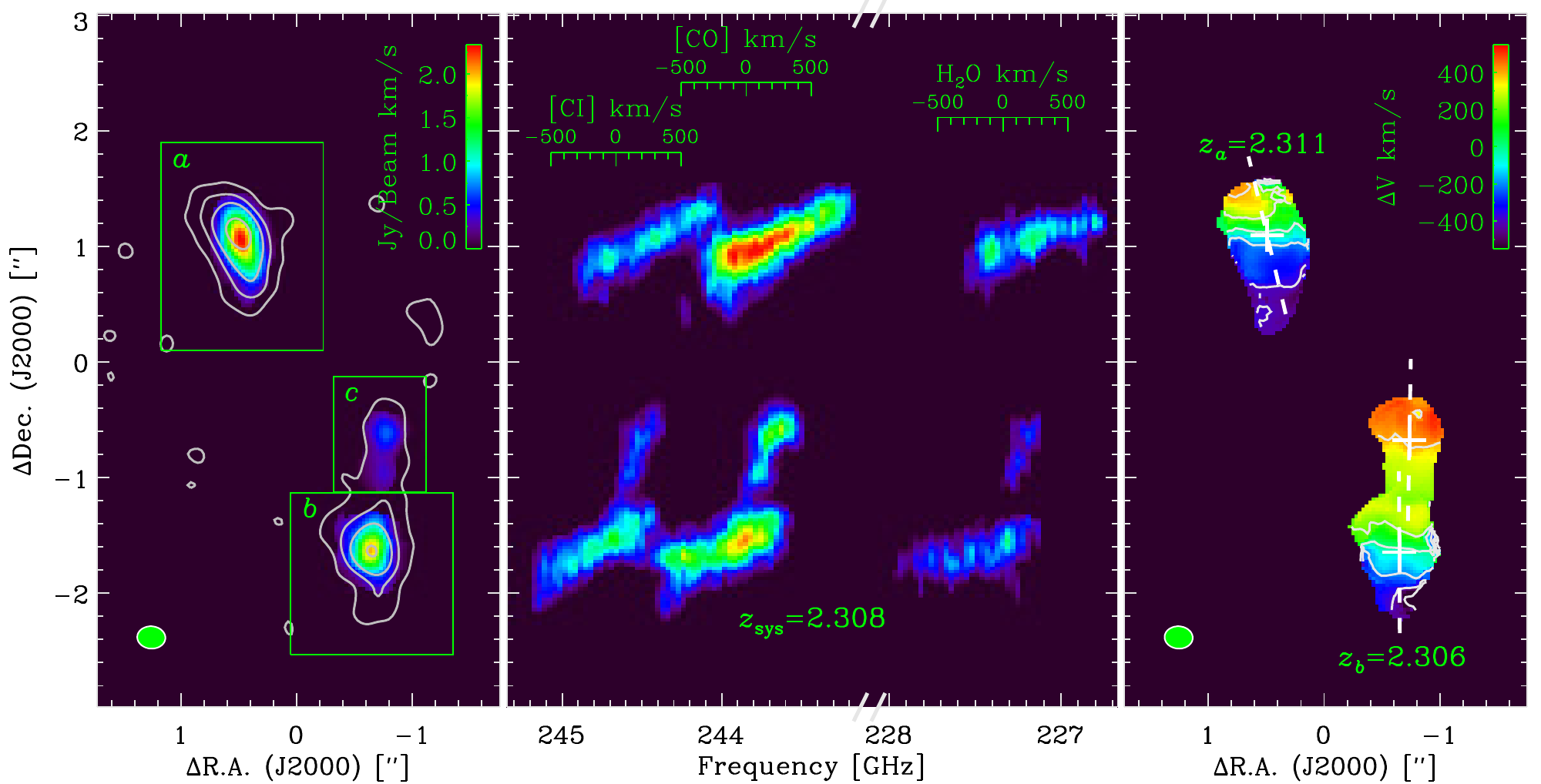}
\caption{
\label{fig:moms}
{\it Left}: Integrated intensity map (color scale) from the combined flux of \citwo, \coseven, and \water\ lines, overlaid with the 230\,GHz continuum (gray contours). The green boxes delineate the apertures adopted for components $a$, $b$, and $c$ (see Table\,\ref{tb:prop}). 
The positional offsets are calculated with respect to $\alpha_{\rm J2000}=02^{\rm h}20^{\rm m}16\farcs613$ and $\delta_{\rm J2000}=-06\arcdeg01\arcmin43\farcs15$.
{\it Middle}: The position-velocity plot generated by collapsing the spectral cubes along the declination dimension. The line velocity scales (computed against the systemic redshift of $z_{\rm sys}$=2.308) are shown near the top.
{\it Right}: The moment-1 image from the \citwo\ and \coseven\ combined spectral cube. The dashed lines show the major axis direction derived from our kinematic disk modeling, and the velocity shift is presented with respect to the systemic redshift of each component.
The synthesized beam is indicated by the green ellipses in the lower-left corners of the moment maps.
}
\end{figure*}

We present the ALMA moment-0/1 maps and the position-velocity (PV) plot in Figure\,\ref{fig:moms}. The moment maps were generated after applying 3D detection masks to the spectral cubes. The masking algorithm first searches for  $\geqslant4\sigma$ continuous regions in 3D smoothed datacubes, then expands each of them to the surrounding 2$\sigma$ contour, and finally pad the regions with an additional 2 pixels in all dimensions. Implementing the masks improves the S/N of moment maps by removing noisy pixels that would overwhelm weak line emission when collapsing the cube along any dimension.

We made clear detection of all of the three targeted lines and spatially resolved \obj\ into three distinct components (labeled as $a$, $b$, and $c$ in Figure\,\ref{fig:moms}). 
In the previous arcsec-resolution \coone\ and dust maps, \obj\ was resolved into a northern and a southern complex \citepalias[see][dubbed as X01N and X01S]{Fu:2013hz}. In the new ALMA data, X01N ($\equiv a$) is clearly elongated along the direction of P.A.$\approx$10\arcdeg, and X01S is further resolved into two separate components ($\equiv b + c$).
Furthermore, all spatially distinct components show systematic velocity gradients in all three lines, with kinematic major axes almost aligned with one another.

While different lines show similar velocity gradients, the PV plot shows dramatically different brightness distributions along the velocity dimension.
Specifically, the \coseven\ and \water\ emission are more asymmetric than \citwo, and are dominated by a few prominent clumps. 
This difference is somewhat expected, because the high critical densities and excitation temperatures of \coseven\ and \water\ lines (see Table\,\ref{tb:lines}) make them great tracers of hot dense clouds \citep[e.g.,][]{Liu:2017ce,Omont:2013kc}, while the lower critical density and excitation temperature of \citwo\ makes it an ideal tracer of neutral gas in moderate physical conditions. The three tracers thus complement one another.  

\begin{figure}\centering
\includegraphics[width=0.48\textwidth]{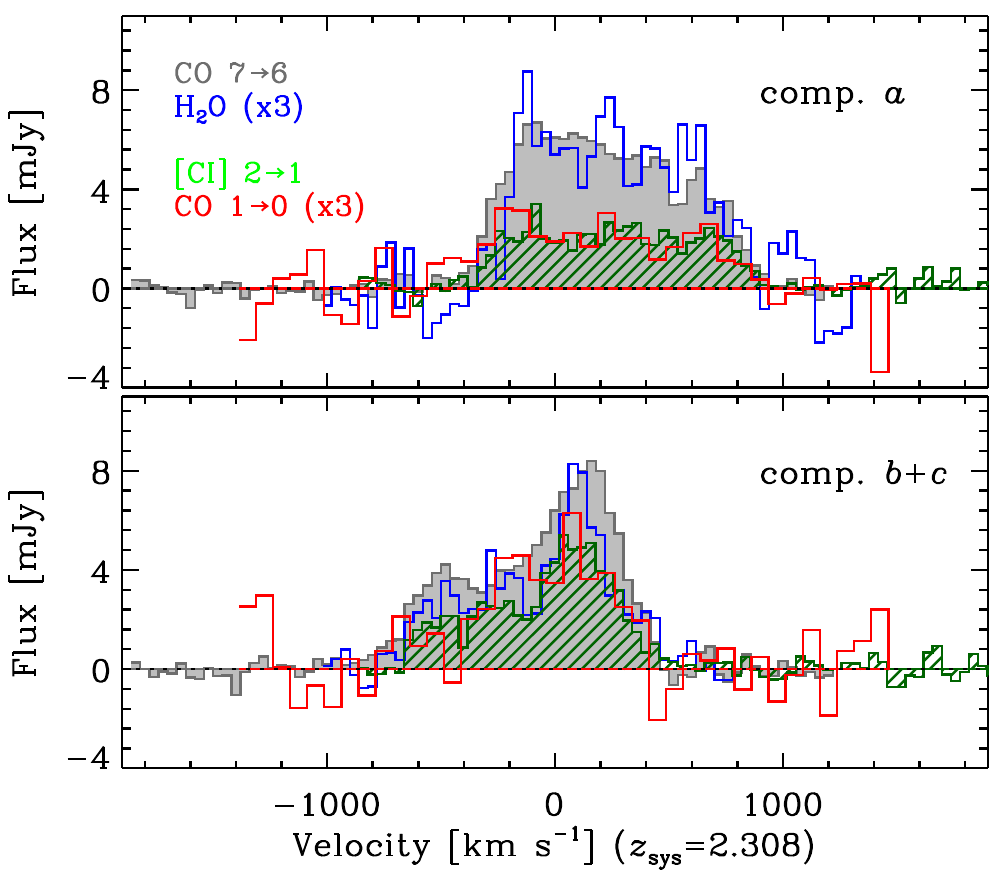}
\caption{
\label{fig:spec}
Spatially-integrated spectra of \citwo, \coseven, \water, and \coone\ towards X01N ($\equiv a$) and X01S ($\equiv b + c$). The \coone\ and \water\ spectra have been scaled up by a factor of 3 in brightness. The \coseven\ and \citwo\ emission are separated in the frequency space by applying a blanking mask before integration.}
\end{figure}

We compare the integrated spectra of different transitions in Figure\,\ref{fig:spec} and present the velocity-integrated line fluxes in Table\,\ref{tb:prop}. The spectra are extracted from the rectangular apertures illustrated in Figure\,\ref{fig:moms}. 
We combine the spectra from components $b$ and $c$ because they are unresolved in \coone. Although the inclusion of $c$ contributes partly to the emission excess at velocities greater than zero \kms, the asymmetric \coseven\ line profiles in both component $a$ and $b$ clearly results from non-axisymmetric surface brightness distributions (as can be seen in Figure\,\ref{fig:moms}).
As expected from their similar critical densities, the line profiles of \coone\ and \citwo\ are in excellent agreement, and a similar level of agreement is observed between \coseven\ and \water.
Interestingly, although the \coone, \citwo, and continuum brightness are comparable between component $a$ and $b + c$ (see Table\,\ref{tb:prop}), component $a$ clearly exhibits a stronger level of \coseven\ and \water\ emission, which may suggest a higher fraction of hot molecular gas.

With intrinsic luminosities of $L_{\rm H_{2}O\,{2_{11}}\to{2_{02}}}\simeq2\times10^8$\,\lsun\ and $L_{\rm IR}\simeq2\times10^{13}$\,\lsun, HXMM01 falls right on the H$_2$O-IR luminosity relation found in local/high-$z$ ultra-luminous IR galaxies (ULIRGs, $10^{12}\lsun<\lir<10^{13}\lsun$) and HyLIRGs \citep{Omont:2013kc,Yang:2013dj,Yang:2016kr}.
By incorporating the lower-$J$ CO flux measurements from \citetalias[][]{Fu:2013hz} with the new \coseven\ measurements, we obtain the global CO spectral line energy distribution (SLED) for HXMM01. 
The CO SLED shape resembles those found in local ULIRGs \citep{Weiss:2005fk,Rangwala:2011dd}, in a sharp contrast with the result of the Milky Way \citep{Fixsen:1999dw} or the high-redshift BzK galaxy samples \citep{Daddi:2015eq}, with significantly higher fraction of CO luminosity distributed at higher-$J$ transitions.
Despite the $S_{\rm\,CO\,7\to6}$/$S_{\rm\,CO\,1\to0}$ brightness ratio decreases by about half from component $a$ to $b$ (from $\simeq9$ to $\simeq4-5$), it is still significantly higher the Galactic center value of $\sim0.9$.
A non-LTE radiative transfer modeling analysis using \radex\ \citep{vanderTak:2007be} suggests that the SLED can be fitted with a two-component model: a low-excitation gas component with $n_{\rm H_{2}}=10^{3.8}-10^{4.7}$\,cm$^{-3}$ and $T_{\rm kin}=15-33$\,K; a high-excitation one ($n_{\rm H_{2}}=10^{2.9}-10^{3.6}$\,cm$^{-3}$ and $T_{\rm kin}=55-132$\,K), which is likely associated with intense on-going star formation.

\subsection{Kinematic Modeling with Tilted-Ring Models}
\label{sec:kinematics}
\label{sec:modelbasic}
\label{sec:fitmethod}
\label{sec:fitpar}

Our observational results reveal that all of the three resolved components are elongated and their light distribution major axes align with monotonic velocity gradients, both of which are indicators of disk-like structures \citep[e.g.,][]{Wisnioski:2015gk,ForsterSchreiber:2018uq}. 
Although we do not have detections of a typical ``spider'' diagram \citep{vanderKruit:1978cn}, it is not likely expected in high-inclination and moderately-resolved disks.
Based on these observational results, we expect that each component is likely well described by a disk-like structure (hereafter simply ``disk''). Therefore, we decide to extract gas kinematics from the ALMA data cubes with tilted-ring models and a Monte-Carlo Markov Chain (MCMC) sampler.

We use the tilted-ring modeling code \tr\footnote{\url{http://gigjozsa.github.io/tirific/}} \citep{Jozsa:2007ey} to simulate spectroscopic cubes. By comparing our data with the simulated spectral cubes from \tr, we simultaneously constrain the rotation curves and line surface brightness (SB) distributions. 
The adopted 3D modeling approach has three major advantages. First, compared with the 2D velocity-field methods, 
we do not fit the extracted velocity fields, which are severely affected by beam smearing\footnote{The beam smearing effect may artificially inflate the observed gas dispersion and reduce the measurable velocity gradient, because it can combine line emission from regions with different radial velocities into a single spectrum.}
in marginally resolved observations of high-redshift galaxies \citep[e.g., see][]{Davies:2011jy}. Instead, the model output is a synthetic spectral cube that includes observational effects such as beam searing and instrumental spectral smoothing. This forward-modeling approach maximally preserves the integrity of the data.
Secondly, we can generate synthetic cubes that include multiple spatial and kinematic components. Thus, we can avoid object or line deblending before modeling.
Finally, \tr\ allows distortions to the SB distribution within rings and can model non-axisymmetric features that are evident in our data.

Specifically, we fit each component to a parametrized rotating disk model, which consists of multiple concentric rings. The disk geometry is described by its center position, its inclination angle from the line-of-sight ($i$), and the P.A. of the projected major axis, all of which are fixed to be the same for different rings.
The kinematics are described by the systemic velocity, the radius-dependent rotational velocity (i.e. rotation curve), and the isotropic velocity dispersion.
In practice, the rotation curve is parameterized by a set of circular velocities on a grid of ring radii.
We adopt a step size of 0\farcs1, which is roughly half of the beam FWHM. The tilted-ring model is evaluated using a spline-interpolated smooth rotation curve based on the discrete circular velocities.
In our case, we assume that all observed lines follow the same rotation curve and their differences in the datacubes are due to different SB distributions and line-of-sight velocity dispersion.

In previous high-$z$ studies, there is also no clear evidence for a systematically varying velocity dispersion as a function of galactocentric distance \citep{DiTeodoro:2016hf,Genzel:2017fa}. Therefore, for simplicity we assume constant velocity dispersion across each disk but allow it to vary among different lines.
Due to the relatively large step size (0\farcs1 = 0.84\,kpc), the velocity dispersion in the model inevitably contains both the cloud-scale gas turbulence and the kpc-scale velocity shear (see further discussion  in \S\,\ref{sec:discussion}), making it difficult to study its radial-dependence robustly. 

To properly model the emission radial profile and asymmetry that is shown in our data (Figure\,\ref{fig:moms}), we adopt a radial- and azimuth-dependent SB prescription for each line. We assume the averaged ring SB follows an exponential intensity profile: $I=I_{0}\exp(-R/r_s)$. The SB variation within each ring is modeled by a first-order sinusoidal distortion, characterized by its amplitude and the node angle relative to the approaching-side major axis. We fix the node angle for all rings to reduce the number of free parameters.

To find the best-fit model and to estimate its uncertainty, we use the Python Affine Invariant MCMC Ensemble sampler \emcee\footnote{\url{http://dfm.io/emcee}} \citep{Goodman:2010et}.
We define the likelihood function of a model given the data as,
\begin{eqnarray}
\label{eq:likelihood}
\ln p & = & -\frac{1}{2}\sum\limits_{i}\left[ \frac{(I_i-M_i)^2}{s_i^2} + \ln(2\pi s_i^2)   \right].
\end{eqnarray}
Here, $\ln p$ is the log-likelihood function, $I_i$ and $M_i$ are the specific intensity of $i$-th pixel in the resampled data and model (see below), respectively, and $s_i=\eta\sigma_{i}$ is the estimator of the ``true'' data uncertainty, with $\sigma_{i}$ from the uncertainty cube of our imaging products.
The scaling factor $\eta$ (close to unity) is introduced to correct uncertainty under/overestimation, and will be precisely determined by the MCMC analysis.

In Equation\,\ref{eq:likelihood}, individual data errors are assumed Gaussian and independent.
To exclude the significant covariance between adjacent pixels in the datacube, we only consider independent beam elements when evaluating the likelihood function. 
To choose independent beam elements, we create a tilted hexagon pattern,
in which a beam FWHM ellipse inscribes each hexagon element.
Then, we extract the 1D spectrum at the center of each element by performing a linear interpolation on the datacube in two spatial dimensions. 
Only these ``resampled'' spectra are used in the likelihood evaluation.
Starting from bounded ``flat'' priors for all free parameters, we iterate with \emcee\ until the posterior distribution of model parameters are sampled adequately.
The posterior distribution functions provide the confidence intervals of the model parameters.

Because of the frequency proximity of \citwo\ and \coseven\ ($\Delta v \approx 1,000$\,\kms\ in the velocity frame), we combine their models into a single simulated spectral cube and compare it with the data cube imaged from two partially overlapping 2\,GHz spectral windows. We simultaneously model the two lines to better constrain the model. The \water\ line is not included in the kinematic modeling due to low S/N.

\begin{figure*}\centering
\includegraphics[width=0.95\textwidth]{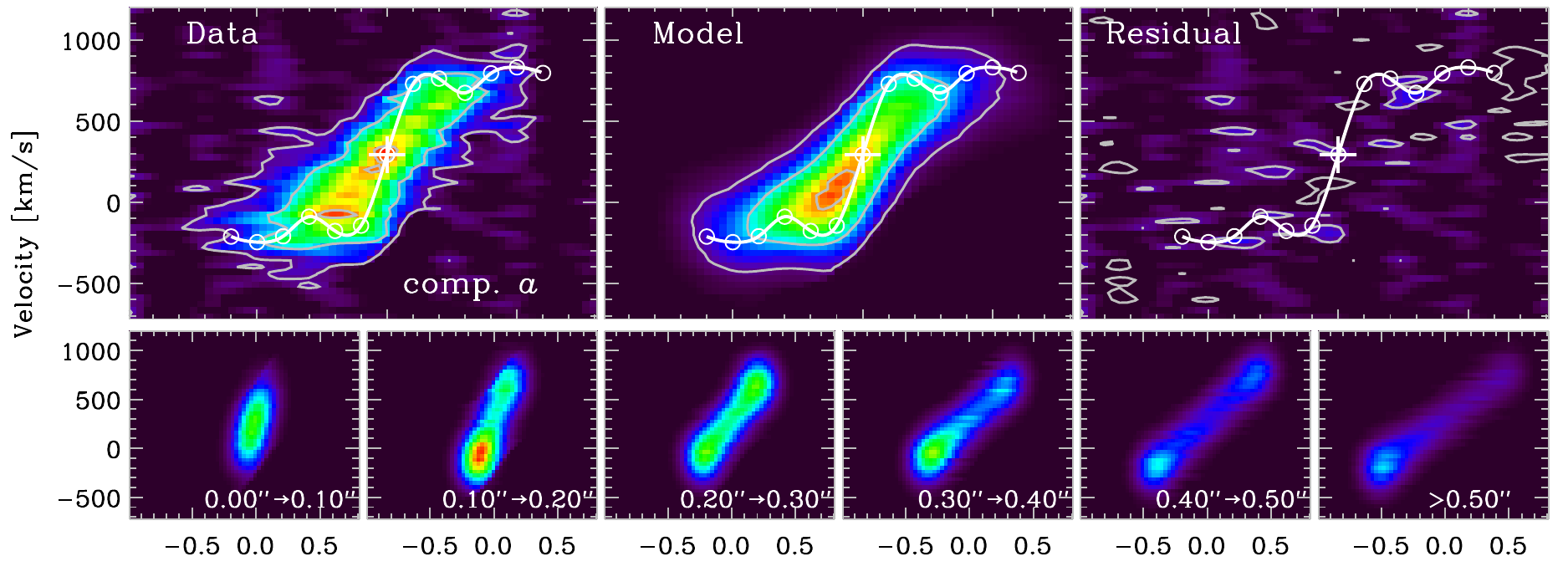}
\includegraphics[width=0.95\textwidth]{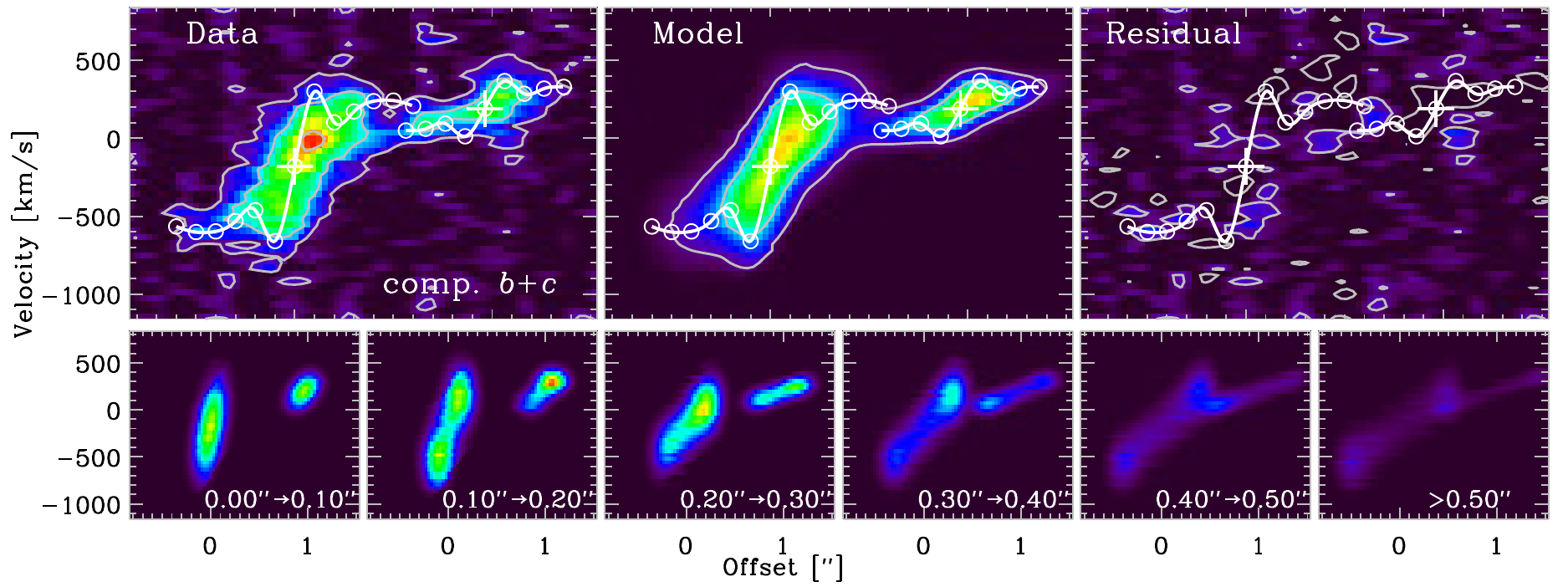}
\caption{\label{fig:pvmom0}
Modeling results of X01N (component $a$) and X01S (component $b+c$), showing as the PV moment-0 maps, by collapsing and summing \coseven\ and \citwo\ emissions along the best-fit minor axes of components $a$ and $b$, respectively. 
The x-axis represents the spatial offset along the major axis of the disk model center for component $a$ or $b$, and the y-axis is in the line-of-sight radial velocity calculated at $z_{\rm sys}=2.308$.
Larger panels show the data, models, and residuals, with contours representing 10, 40, and 90\% of the peak values from the data map. The disk rotation curves are overlaid in white lines after the amplitudes are multiplied by inclination correction factors of $\sin(i)$.
The smaller panels below show the model emission contributed by individual disk rings, labelled with their radii.
}
\end{figure*}

To show the quality of the best-fit models, we compare the PV maps from the \citwo\ and \coseven\ data cube and the best-fit \tr\ models in Figure\,\ref{fig:pvmom0}. We also illustrate the contributions from individual rings to the PV maps. We report the best-fit parameters and their uncertainties in Table\,\ref{tb:fitres_geo}. The rotation curves of the two major components ($a$ and $b$) are plotted in Figure\,\ref{fig:radpro_ns}. We discuss the modeling results in the next subsection.

\subsection{Rotation Curve, Dynamical Mass, and \aco}
\label{sec:ModelRes}

\begin{figure*}\centering
\includegraphics[width=0.496\textwidth]{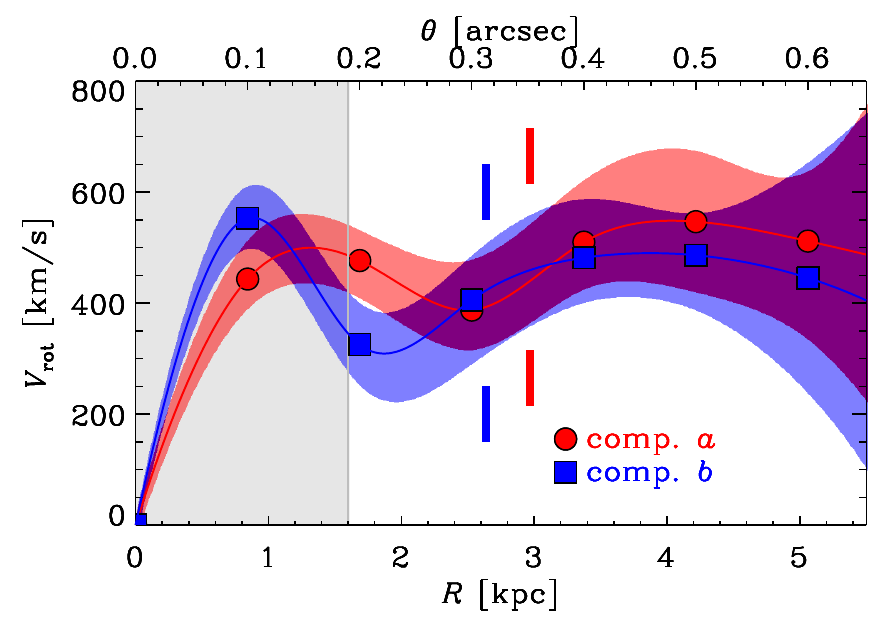}
\includegraphics[width=0.496\textwidth]{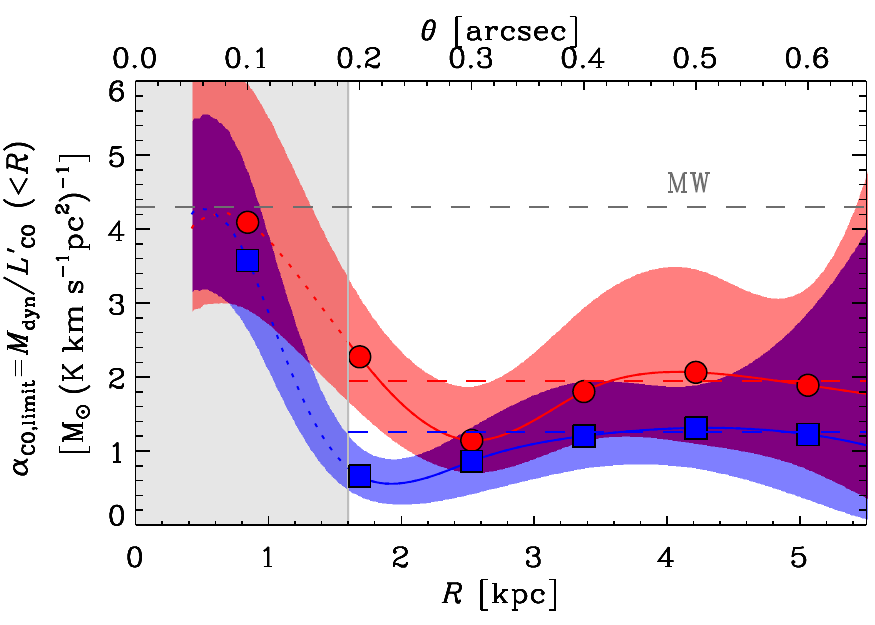}
\caption{\label{fig:radpro_ns}\label{fig:radpro_alpha}
{\it Left:} Inclination-corrected rotation curves of components $a$ and $b$, with the uncertainties (defined by the 2.5th and 97.5th percentiles) depicted with the shaded areas.
The vertical bars show the half-light radii expected from the exponential gas disk models ($r_{\rm 1/2}=1.678{r_{\rm s, CI}}$).
{\it Right:} Upper limits of $\alpha_{\rm CO}$ vs. galactocentric distance. The upper limits are derived from the ratio of enclosed dynamical mass (assuming thin disk mass distribution) and \coone\ luminosity (based on exponential models derived from the \citwo\ map). The dashed red and blue lines show the median values of $\alpha_{\rm CO}$ limits between 3 and 5\,kpc.
The canonical Galactic $\alpha_{\rm CO}$ value is shown in the dashed gray line.
In both panels, the gray boxes present the beam FWHM, therefore, the results are likely dependent on depending on some model presumptions (e.g. radial-independent gas dispersion) and numerical interpolations.
}
\end{figure*}

\begin{deluxetable*}{ccccc}[htb!]
\tablecolumns{3}
\tablewidth{0pt}
\tablecaption{Properties and Modeling Results by Regions \label{tb:prop}\label{tb:fitres_geo}}
\tablehead{ 
\colhead{Quantity} &
\colhead{Unit} &
\colhead{Comp. $a$} & 
\colhead{Comp. $b$} &
\colhead{Comp. $c$}}
\startdata
\toprule
\multicolumn{5}{c}{Observed Properties} \\
\midrule
$S_{\rm\,CO\,7\to6}$				& Jy\,\kms 											& $6.2\pm0.2$ 						& $3.5\pm0.1$ 						& $0.9\pm0.1$ 		\\
$S_{\rm\,[CI]\,2\to1}$ 				& Jy\,\kms 											& $2.7\pm0.2$ 						& $2.6\pm0.1$ 						& $0.5\pm0.1$  	\\
$S_{\rm\,H_{2}O}$ 	& Jy\,\kms 														& $2.4\pm0.2$ 						& $1.1\pm0.1$ 						& $0.3\pm0.1$ 		\\
$S_{\rm\,CO\,1\to0}$ 				&  Jy\,\kms 											& $0.7\pm0.1$ 						& \multicolumn{2}{c}{$0.9\pm0.2$}						\\
$S_{\rm\,230\,GHz}$ 				&  mJy 												& $4.3\pm0.1$  					& $3.6\pm0.1$ 						& $0.5\pm0.1$ 		\\
\midrule
\multicolumn{5}{c}{Dynamical Models - Kinematics} \\
\midrule
$\alpha_{\rm J2000}$ 			& hh:mm:ss.sss										&02:20:16.653						& 02:20:16.576						& 02:20:16.569	\\
$\delta_{\rm J2000}$ 			& dd:mm:ss.ss											&-06:01:41.92						& -06:01:44.60 						& -06:01:43.62	\\
$V_{\rm sys}$					& \kms 												& $+292_{-21}^{+26}$				& $-179_{-22}^{+24}$				& $+189_{-27}^{+30}$ \\ 
P.A.							& degree 												&$+14\pm3$						& $+1\pm3$ 						& $-2\pm4$	\\
$i$							& degree 												&$79\pm2$						& $60\pm5$ 						& $69\pm4$	\\
$\sigma_{{\rm CO\,7\to6}}$ 		& \kms 												&$156_{-14}^{+16}$					& $168_{-24}^{+17}$ 				& $\phn66_{-11}^{+14}$	\\
$\sigma_{{\rm CI}}$				& \kms 												&$150_{-26}^{+24}$					& $166_{-28}^{+22}$ 				& $\phn65_{-26}^{+24}$	\\
$V_{{\rm max}}$				& \kms 												&$542_{-108}^{+105}$				& $486_{-81}^{+54}$					& $168_{-72}^{+82}$		\\
\midrule
\multicolumn{5}{c}{Dynamical Models - Surface Brightness Distribution} \\
\midrule
$\mathcal{S}_{{\rm\,CO\,7\to6}}$ 		& Jy\,\kms											&$5.8_{-0.2}^{+0.4}$					&$3.0_{-0.2}^{+0.2}$					&$1.3_{-0.2}^{+0.1}$	\\
$\mathcal{S}_{\rm\,[CI]\,2\to1}$					& Jy\,\kms									&$3.0_{-0.2}^{+0.3}$					&$2.9_{-0.2}^{+0.3}$					&$0.8_{-0.2}^{+0.3}$	\\
$\mathcal{S}_{\rm\,CO\,1\to0}$  		& Jy\,\kms 										&$0.6_{-0.1}^{+0.2}$					&$0.6_{-0.1}^{+0.2}$					&$<0.2$	\\
$r_{\rm s,CO\,7\to6}$ 		& kpc 													&$1.6_{-0.1}^{+0.1}$					&$1.2_{-0.1}^{+0.1}$					&$1.1_{-0.2}^{+0.1}$	\\
$r_{\rm s,CI}$ 					& kpc 												&$1.8_{-0.1}^{+0.1}$					&$1.6_{-0.1}^{+0.1}$					&$1.5_{-0.3}^{+0.1}$	\\
$r_{\rm s,CO\,1\to0}$ 		& kpc 													&$2.1_{-0.9}^{+0.5}$					&$2.5_{-1.0}^{+0.9}$					& \nodata	\\
\midrule
$M_{\rm dyn,e}^{\rm 5\,kpc}$			& $10^{11}\,{M_{\odot}}$												&$2.3_{-0.5}^{+0.3}$			& $1.8_{-0.5}^{+0.6}$ 			& \nodata	\\
$\alpha_{\rm CO,limit}$			& ${M_{\odot}}$$(\Kkmsec {\rm pc}^{2})^{-1}$					&$2.0_{-0.6}^{+0.7}$					& $1.4_{-0.5}^{+0.4}$ 				& \nodata	\\
\bottomrule
\enddata
\tablecomments{The velocity was computed against a systemic redshift of $z_{\rm sys}=2.308$. The line/continuum flux is measured within individual rectangular apertures defined in Figure\,\ref{fig:moms}.
The $\alpha_{\rm CO}$ limits are the median values estimated within a radius range between 3 and 5\,kpc (see Figure\,\ref{fig:radpro_alpha}).
In the \coone\ modeling, the scale-length for component $c$ is set to be the same as $\mathcal{S}_{\rm\,[CI]\,2\to1}$.}
\end{deluxetable*}

The kinematic models show that components $a$ and $b$ exhibit similar rotation curves: the circular velocity rises rapidly within 1\,kpc and eventually reaches a plateau of $\sim500$\,\kms. Beyond $\sim5$\,kpc, the S/N of the line emission becomes too low to trace the kinematics.
All disk models suggest $V_{\rm max}/\sigma_{0}\gtrsim3$, where $\sigma_{0}$ is the intrinsic gas dispersion\footnote{The values of $\sigma_{{\rm CO\,7\to6}}$ and $\sigma_{{\rm CI}}$ are estimated from dynamical modeling, under the assumptions detailed in \S\,\ref{sec:kinematics}. They likely only provide upper limits for the intrinsic gas dispersion $\sigma_{0}$ due to kinematic structures below the resolution limit.
This indicates the systems are mainly rotationally supported rather than pressure supported, providing further evidence to support the assumption of ``disk-like'' structures.}

Assuming that each component is {\it entirely} supported by ordered rotation, we can estimate the dynamical mass as a function of radius using the best-fit rotation curves.
For a spherically symmetric distribution, the dynamical mass is simply,
\begin{eqnarray}\label{eq:dmass1}
M_{\rm dyn,s}=\frac{V_{\rm rot}^2R}{G}.
\end{eqnarray}
For a geometrically thin disk with an exponential mass distribution, i.e, $\Sigma(R)\propto\exp(-R/r_{\rm s,mass})$, the enclosed dynamical mass within $R$ is,
\begin{eqnarray}\label{eq:dmass2}
M_{\rm dyn,e}=\frac{V_{\rm rot}^2R}{G}\frac{1-e^{2y}\left(1+2y\right)}{4y^3[I_0(y)K_0(y)-I_1(y)K_1(y)]},
\end{eqnarray}
where $y\equiv R/(2r_{\rm s,mass})$, and $I_i$ and $K_i$ are the modified Bessel functions \citep[see][\S\,2.6]{Binney:2008wd}. 
While Equation\,\ref{eq:dmass1} is suitable for the scenario where the mass is dominated by a dark matter (DM) halo or a stellar bulge, we adopt Equation\,\ref{eq:dmass2} because the gravitational potential is likely dominated by a gas-rich disk in our case.

An enclosed dynamical mass gives an upper limit on the molecular gas mass. 
The ratio between the dynamic mass and the corresponding integrated \coone\ line flux provides strict upper limit of the \cohh\ conversion factor $\alpha_{\rm CO}$.
Assuming that the scale-lengths of the disk mass and \coone\ brightness distribution ($r_{\rm s,mass}$ and $r_{\rm s,CO\,1\to0}$, respectively) are the same as that of the \citwo, we calculate the enclosed dynamical mass and luminosity-weighted $\alpha_{\rm CO}$ upper limit as a function of radius for each component.
The similar line profiles of \citwo\ and \coone\ indicate the high-resolution morphology of \citwo\ should be a good approximation for the \coone\ distribution, which was marginally resolved in the \coone\ map.

Keeping the model assumptions in mind, the results reveal that the dynamical mass within 5\,kpc reach $\sim2\times10^{11}$\,\msun for both of two major components ($a$ and $b$) (see Table\,\ref{tb:fitres_geo}).
We present the derived $\alpha_{\rm CO}$ upper limit as a function of radius in the right panel of Figure\,\ref{fig:radpro_ns}.
As the radius increases, the accumulation of the line flux roughly cancels out the increase in dynamical mass, producing an approximately constant \cohh\ conversion factor at $3\lesssim R \lesssim5$\,kpc. 
The median $\alpha_{\rm CO}$  limits estimated from this radius range equal 1.4 and 2.0\,${M_{\odot}}$/$\Kkmsec {\rm pc}^{-2}$ for $a$ and $b$, respectively. 
Both values are consistent with the low $\alpha_{\rm CO}$ of $\approx0.6-0.8$\,\unitaco\ found in local ULIRGs \citep[e.g.,][]{Downes:1998co,Papadopoulos:2012hj}.

For an unlikely spherically symmetric mass distribution, the dynamical mass will increase by $10-30$\% within $R=3-5$\,kpc.
Additional dynamical mass and $\alpha_{\rm CO,limit}$ uncertainties can arise from the adoption of \citwo\ scale-length for the disk density and \coone\ brightness profiles. 
We experiment with our 3D modeling approach to evaluate the \coone\ SB profile, using our reprocessed VLA \coone\ datacube and the best-fit kinematic models from the high-resolution ALMA data (\S\,\ref{sec:kinematics}). 
The resulting scale-lengths $r_{\rm s,CO\,1\to0}$ show significant uncertainties due to the moderate resolution and SNR of the VLA map ($2.1_{-0.9}^{+0.5}$\,kpc and $2.5_{-1.0}^{+0.9}$\,kpc, respectively, see Table\,\ref{tb:fitres_geo}).
However, the values do agree with those of \citwo\ within the error margins.
If we explicitly adopt $r_{\rm s,CO\,1\to0}$ for deriving $\alpha_{\rm CO,limit}$ (despite the large error bars), the $\alpha_{\rm CO,limit}$ values will increase to 2.2\,${M_{\odot}}$/$\Kkmsec {\rm pc}^{-2}$ for both components.
Nevertheless, the above estimation still suggest that the $\alpha_{\rm CO}$ value in HXMM01 is lower than the canonical Galactic value of 4.3\,\unitaco\ \citep{Bolatto:2013hl}.

We caution that we do {\em not} include any lensing correction in the analysis of dynamical mass and $\alpha_{\rm CO}$ and the presentations of physical scales, due to the uncertainties in the lensing model.
Assuming that the lensing magnification is the same along the major and the minor axes of each disk, the disk inclination (therefore $V_{\rm max}$) would remain unchanged, but its physical scales and luminosity would be overestimated by a factor of $\sqrt{\mu}$ and $\mu$, respectively, where $\mu$ is the magnification factor.
Taking an average magnification factor of $\mu\approx1.6$ from \citetalias{Fu:2013hz}, a lensing correction may increase our $\alpha_{\rm CO}$ upper limits by up to 30\%.

\section{Discussion}

\label{sec:discussion}
\label{sec:summary}

The derived rotation curves of components $a$ and $b$ do not deviate from the typical one found in local spiral galaxies \citep[e.g.,][]{Rubin:1970gu,Begeman:1989wv}, including the Milky Way, which is characterized by a rapidly rising velocity followed by an extended flat portion \citep{Clemens:1985dp}. However, their rotation curves rarely reach the amplitude in HXMM01 ($\sim500\,\rm km\,s^{-1}$).
The shape suggests a concentration of baryonic mass in the central $\sim2$\,kpc.
This is consistent with the compact morphology of high-$z$ galaxies measured from starlight \citep[e.g.,][]{Bruce:2012dn,vanderWel:2014hi} and gas tracers \citep[e.g.,][]{Tacconi:2008fe,Hodge:2012jt}.
We do not find clear evidence of decreasing rotation velocity in the outer part of each component. Such ``declining'' rotation curves were identified in some previous H$\alpha$-based studies \citep{Genzel:2017fa,Lang:2017bv}, which have been suggested as evidence for either a lack of dark matter or significant gas pressure in these high-$z$ systems.
The discrepancy could be observational or intrinsic:
our data only provide circular velocity out to $\sim4-5$\,kpc, not as far as the radii reached by those studies (up to 10\,kpc);
on the other hand, their sample consists of isolated main-sequence ``normal'' star-forming galaxies, which may exhibit different baryon distributions relative to dark matter halo or show more pronounced pressure-supporting effect in outer disks \citep{Burkert:2010cv}.
We note that \citet[][]{Levy:2018hj} reports lower circular velocity of ionized gas than that of neutral gas in the outer disks of their local disk galaxy sample, likely due to thick and turbulent disk of ionized gas.
Therefore, we may need to take the systematic difference among gas tracers into interpretation. 

One important caveat of interpreting kinematics in HXMM01 is that it is a rare starburst merging system and the interaction among different components might cause non-equilibrium gas motions.
Such tidally induced kinematic disorders are more likely to present in the outskirts of galaxy disks.
Although we cannot to rule out the influence of such interaction based on existing data, we expect it plays a minor role on the gas kinematics at the galactocentric radii we are able to probe, considering the large separation of two major components ($\geq$25\,kpc).
On the other hand, the \coseven\ distribution in two major components is clearly skewed towards their interacting partners, while the morphological asymmetry is almost absent in \citwo (see Figure\,\ref{fig:moms}).
Because \coseven\ traces the high-density warm molecular gas, this could be evidence for elevated star formation efficiency due to galaxy interaction, rather than substantial perturbation to the gas kinematics or mass distribution.

While the outer rotation curves provide critical constraints for the baryon/DM distribution in high-redshift objects, the brightness of tracers generally fall rapidly.
The inaccessibility of the \hone\ 21cm line prompts the search for alternative kinematic tracers of neutral gas.
Based on the ALMA data, we find that both \citwo\ and \coone\ emission are more extended than \coseven\ and \water, as indicated by the larger scale-lengths. This result is consistent with previous multi-transition studies \citep[e.g.,][]{Ivison:2011it,Riechers:2010kf}.
The integrated line profiles and distributions of \citwo\ and \coone\ are strikingly similar, and their line ratios are also consistent among the disks. 
A similar characterization was also found in previous Galactic surveys \citep[e.g.,][]{Ojha:2001gs} and examined via time-dependent chemical modeling \citep{Papadopoulos:2004jl}.
The extended morphology of [\ion{C}{1}], the brightness strength of \coseven, and their close frequency make the pair a complementary tracer combination. 
Similar to the discussion in \citet{Papadopoulos:2004jl}, we believe that this combination is the best surrogate to \hone\ 21cm, low-$J$ CO, or [\ion{C}{2}] for studying gas dynamics in the inner regions and outskirts of high-$z$ star-forming galaxies.

The high IR luminosity in HXMM01 implies a minimum molecular gas mass of $M_{\rm mol} \geq L_{\rm IR}/(500\,L_{\odot}/M_{\odot}) = 4\times10^{10}\,M_{\odot}$, where the maximum light-to-mass ratio is given by the Eddington limit \citep{Scoville:2004uf,Thompson:2005dq}.
Combined with the \coone\ luminosity of $L^{\prime}_{\rm CO} = 3 \times 10^{11}$\,K\,\kms\,pc$^{2}$, we obtain a lower limit on the \cohh\ conversion factor of $\alpha_{\rm CO}\geq0.13$. 
A different lower limit on the $\alpha_{\rm CO}$ can be obtained by assuming local thermodynamic equilibrium (LTE) and an optically thin \coone\ transition \citep{Ivison:2011it}.
The result varies from 0.4 to 0.6\,\unitaco, depending on the adopted gas kinetic temperature ($T_{\rm kin}=15-50$\,K).
On the other hand, our dynamical mass estimation provides a strict upper limit of $\alpha_{\rm CO}\lesssim1.4-2.0$\,\unitaco.
Although both approaches only provide limits, the results conclusively show that $\alpha_{\rm CO}$ in HXMM01 is lower than the canonical Galactic value at least by a factor of 2. This is compatible with other measurements in local or high-$z$ starburst galaxies \citep[e.g.,][]{Downes:1998co,Papadopoulos:2012hj,Hodge:2012jt}.
Unless the SFR in HXMM01 is significantly overestimated \citep[e.g. due to a top-heavy IMF,][]{Zhang:2018ee},
the gas-exhausting timescale is still short at most $\sim200$\,Myr.

As an important dynamical state parameter, the $V_{\rm max}/\sigma_0$ values of componnets $a$ and $b$ reach $\sim$3, lower than those found in $z\sim1-2$ normal star-forming disk galaxies estimated from H$\alpha$ observations \citep{Cresci:2009dv,DiTeodoro:2016hf}.
Their disk galaxy samples have moderate SFR ($\lesssim200$\,\msunyr) and show much lower rotational velocities ($V_{\rm max}\sim100-300$\,\kms).
Their observed gas dispersion is significantly lower ($\sim20-80$\,\kms) than what is required in our best-fit models for components $a$ and $b$ ($>100$\,\kms).
\citet{Burkert:2010cv} discussed a partially pressure-supported disk, in which the radial pressure partly counteracts the gravitational force, reducing the observable gas rotational velocities.
Following their pressure-corrected model, we found that the estimated dynamical mass may increase by at most $\sim50$\%.
On the other hand, the gas dispersion derived from kinematic modeling ($\sigma_{{\rm CO\,7\to6}}$ or $\sigma_{{\rm CI}}$) should be only considered as the upper limit of intrinsic gas velocity dispersion due to unresolved kinematic structures such as sub-kpc scale velocity shear. Even with 3D modeling, the gas kinematics at different spatial scales will still become distinguishable as the data resolution degrades, especially near galactic centers.
By examining the data and best-fit model cubes, we find that our best-fit models of components $a$ and $b$ do overestimate the line widths at large radii ($\gtrsim3$\,kpc) while provide good fit for inner disks.
Therefore, the $V_{\rm max}/\sigma_{0}$ values is likely higher than the ones indicated by our models.
We experiment alternative models by fixing the gas dispersion of all disk rings to the values directly measured from outer-ring line profiles ($\sigma_{\rm outer}\sim60-80$\,\kms\ after instrumental correction). However, the goodness of fit degrades for inner disks.
It is possible that the gas dispersion at smaller galactocentric radii is intrinsically larger, contradicting to our radially constant dispersion assumption.
It is also likely that the disk brightness and dynamical structures are more complex than the prescription adopted in our models.
While higher resolution data are required to distinguish different possibilities, both of dynamical modeling and line profile measurement show that components $a$ and $b$ in HXMM01 are still highly turbulent ($\gtrsim60$\,\kms).

\acknowledgments

R.X., H.F., and J.I. acknowledge support from the National Science Foundation (NSF) grant AST-1614326, the National Aeronautics and Space Administration (NASA) JPL award RSA\#1568087, and funds from the University of Iowa. This work was performed in part at Aspen Center for Physics, which is supported by NSF grant PHY-1607611.

The National Radio Astronomy Observatory is a facility of the National Science Foundation operated under cooperative agreement by Associated Universities, Inc. This paper makes use of the following ALMA data: ADS/JAO.ALMA\#2015.1.00723.S, ADS/JAO.ALMA\#2011.0.00539.S.
ALMA is a partnership of ESO (representing its member states), NSF (USA) and NINS (Japan), together with NRC (Canada), NSC and ASIAA (Taiwan), and KASI (Republic of Korea), in cooperation with the Republic of Chile. The Joint ALMA Observatory is operated by ESO, AUI/NRAO and NAOJ.

We thank the anonymous referee for helpful comments and suggestions, which largely improved the clarity of our results. 

\software{CASA \citep{McMullin:2007ta}, Astropy \citep{AstropyCollaboration:2013cd}, {\tt TiRiFiC} \citep{Jozsa:2007ey} {\tt emcee} \citep{Goodman:2010et}}

\bibliographystyle{aasjournal}

\begin{thebibliography}{}
\expandafter\ifx\csname natexlab\endcsname\relax\def\natexlab#1{#1}\fi
\providecommand{\url}[1]{\href{#1}{#1}}
\providecommand{\dodoi}[1]{doi:~\href{http://doi.org/#1}{\nolinkurl{#1}}}
\providecommand{\doeprint}[1]{\href{http://ascl.net/#1}{\nolinkurl{http://ascl.net/#1}}}
\providecommand{\doarXiv}[1]{\href{https://arxiv.org/abs/#1}{\nolinkurl{https://arxiv.org/abs/#1}}}

\bibitem[{{Astropy Collaboration} {et~al.}(2013){Astropy Collaboration},
  Robitaille, Tollerud, Greenfield, Droettboom, Bray, Aldcroft, Davis,
  Ginsburg, Price-Whelan, Kerzendorf, Conley, Crighton, Barbary, Muna,
  Ferguson, Grollier, Parikh, Nair, Unther, Deil, Woillez, Conseil, Kramer,
  Turner, Singer, Fox, Weaver, Zabalza, Edwards, Azalee~Bostroem, Burke, Casey,
  Crawford, Dencheva, Ely, Jenness, Labrie, Lim, Pierfederici, Pontzen, Ptak,
  Refsdal, Servillat, \& Streicher}]{AstropyCollaboration:2013cd}
{Astropy Collaboration}, Robitaille, T.~P., Tollerud, E.~J., {et~al.} 2013,
  A{\&}A, 558, 33

\bibitem[{Barger {et~al.}(1998)Barger, Cowie, Sanders, Fulton, Taniguchi, Sato,
  Kawara, \& Okuda}]{Barger:1998em}
Barger, A.~J., Cowie, L.~L., Sanders, D.~B., {et~al.} 1998, Nature, 394, 248

\bibitem[{Begeman(1989)}]{Begeman:1989wv}
Begeman, K.~G. 1989, A{\&}A, 223, 47

\bibitem[{Binney \& Tremaine(2008)}]{Binney:2008wd}
Binney, J., \& Tremaine, S. 2008, Galactic Dynamics: Second Edition, by James
  Binney and Scott Tremaine. ISBN 978-0-691-13026-2 (HB). Published by
  Princeton University Press, Princeton, NJ USA, 2008.

\bibitem[{Bolatto {et~al.}(2013)Bolatto, Wolfire, \& Leroy}]{Bolatto:2013hl}
Bolatto, A.~D., Wolfire, M., \& Leroy, A.~K. 2013, ARA{\&}A, 51, 207

\bibitem[{Bothwell {et~al.}(2013)Bothwell, Aguirre, Chapman, Marrone, Vieira,
  Ashby, Aravena, Benson, Bock, Bradford, Brodwin, Carlstrom, Crawford,
  De~Breuck, Downes, Fassnacht, Gonzalez, Greve, Gullberg, Hezaveh, Holder,
  Holzapfel, Ibar, Ivison, Kamenetzky, Keisler, Lupu, Ma, Malkan, McIntyre,
  Murphy, Nguyen, Reichardt, Rosenman, Spilker, Stalder, Stark, Strandet,
  Vernet, Weiss, \& Welikala}]{Bothwell:2013ed}
Bothwell, M.~S., Aguirre, J.~E., Chapman, S.~C., {et~al.} 2013, ApJ, 779, 67

\bibitem[{Bruce {et~al.}(2012)Bruce, Dunlop, Cirasuolo, McLure, Targett, Bell,
  Croton, Dekel, Faber, Ferguson, Grogin, Kocevski, Koekemoer, Koo, Lai, Lotz,
  McGrath, Newman, \& van~der Wel}]{Bruce:2012dn}
Bruce, V.~A., Dunlop, J.~S., Cirasuolo, M., {et~al.} 2012, MNRAS, 427, 1666

\bibitem[{Burkert {et~al.}(2010)Burkert, Genzel, Bouch{\'e}, Cresci, Khochfar,
  Sommer-Larsen, Sternberg, Naab, F{\"o}rster~Schreiber, Tacconi, Shapiro,
  Hicks, Lutz, Davies, Buschkamp, \& Genel}]{Burkert:2010cv}
Burkert, A., Genzel, R., Bouch{\'e}, N., {et~al.} 2010, ApJ, 725, 2324

\bibitem[{Bussmann {et~al.}(2015)Bussmann, Riechers, Fialkov, Scudder, Hayward,
  Cowley, Bock, Calanog, Chapman, Cooray, De~Bernardis, Farrah, Fu, Gavazzi,
  Hopwood, Ivison, Jarvis, Lacey, Loeb, Oliver, Perez-Fournon, Rigopoulou,
  Roseboom, Scott, Smith, Vieira, Wang, \& Wardlow}]{Bussmann:2015jx}
Bussmann, R.~S., Riechers, D., Fialkov, A., {et~al.} 2015, ApJ, 812, 43

\bibitem[{Calanog {et~al.}(2014)Calanog, Fu, Cooray, Wardlow, Ma, Amber, Baker,
  Baes, Bock, Bourne, Bussmann, Casey, Chapman, Clements, Conley, Dannerbauer,
  de~Zotti, Dunne, Dye, Eales, Farrah, Furlanetto, Harris, Ivison, Kim, Maddox,
  Magdis, Messias, Michalowski, Negrello, Nightingale, O'Bryan, Oliver,
  Riechers, Scott, Serjeant, Simpson, Smith, Timmons, Thacker, Valiante, \&
  Vieira}]{Calanog:2014ep}
Calanog, J.~A., Fu, H., Cooray, A., {et~al.} 2014, ApJ, 797, 138

\bibitem[{Clemens(1985)}]{Clemens:1985dp}
Clemens, D.~P. 1985, ApJ, 295, 422

\bibitem[{Cresci {et~al.}(2009)Cresci, Hicks, Genzel, Schreiber, Davies,
  Bouch{\'e}, Buschkamp, Genel, Shapiro, Tacconi, Sommer-Larsen, Burkert,
  Eisenhauer, Gerhard, Lutz, Naab, Sternberg, Cimatti, Daddi, Erb, Kurk, Lilly,
  Renzini, Shapley, Steidel, \& Caputi}]{Cresci:2009dv}
Cresci, G., Hicks, E. K.~S., Genzel, R., {et~al.} 2009, ApJ, 697, 115

\bibitem[{Daddi {et~al.}(2015)Daddi, Dannerbauer, Liu, Aravena, Bournaud,
  Walter, Riechers, Magdis, Sargent, B{\'e}thermin, Carilli, Cibinel,
  Dickinson, Elbaz, Gao, Gobat, Hodge, \& Krips}]{Daddi:2015eq}
Daddi, E., Dannerbauer, H., Liu, D., {et~al.} 2015, A{\&}A, 577, A46

\bibitem[{Davies {et~al.}(2011)Davies, Schreiber, Cresci, Genzel, Bouch{\'e},
  Burkert, Buschkamp, Genel, Hicks, Kurk, Lutz, Newman, Shapiro, Sternberg,
  Tacconi, \& Wuyts}]{Davies:2011jy}
Davies, R., Schreiber, N. M.~F., Cresci, G., {et~al.} 2011, ApJ, 741, 69

\bibitem[{de~Blok {et~al.}(2008)de~Blok, Walter, Brinks, Trachternach, Oh, \&
  Kennicutt}]{deBlok:2008cn}
de~Blok, W. J.~G., Walter, F., Brinks, E., {et~al.} 2008, AJ, 136, 2648

\bibitem[{Di~Teodoro {et~al.}(2016)Di~Teodoro, Fraternali, \&
  Miller}]{DiTeodoro:2016hf}
Di~Teodoro, E.~M., Fraternali, F., \& Miller, S.~H. 2016, A{\&}A, 594, A77

\bibitem[{Downes \& Solomon(1998)}]{Downes:1998co}
Downes, D., \& Solomon, P.~M. 1998, ApJ, 507, 615

\bibitem[{Fixsen {et~al.}(1999)Fixsen, Bennett, \& Mather}]{Fixsen:1999dw}
Fixsen, D.~J., Bennett, C.~L., \& Mather, J.~C. 1999, ApJ, 526, 207

\bibitem[{F{\"o}rster~Schreiber {et~al.}(2018)F{\"o}rster~Schreiber, Renzini,
  Mancini, Genzel, Bouch{\'e}, Cresci, Hicks, Lilly, Peng, Burkert, Carollo,
  Cimatti, Daddi, Davies, Genel, Kurk, Lang, Lutz, Mainieri, McCracken,
  Mignoli, Naab, Oesch, Pozzetti, Scodeggio, Shapiro~Griffin, Shapley,
  Sternberg, Tacchella, Tacconi, Wuyts, \& Zamorani}]{ForsterSchreiber:2018uq}
F{\"o}rster~Schreiber, N.~M., Renzini, A., Mancini, C., {et~al.} 2018, eprint
  arXiv:1802.07276

\bibitem[{Fu {et~al.}(2012)Fu, Jullo, Cooray, Bussmann, Ivison, Perez-Fournon,
  Djorgovski, Scoville, Yan, Riechers, Aguirre, Auld, Baes, Baker, Bradford,
  Cava, Clements, Dannerbauer, Dariush, de~Zotti, Dole, Dunne, Dye, Eales,
  Frayer, Gavazzi, Gurwell, Harris, Herranz, Hopwood, Hoyos, Ibar, Jarvis, Kim,
  Leeuw, Lupu, Maddox, Mart{\'\i}nez-Navajas, Michalowski, Negrello, Omont,
  Rosenman, Scott, Serjeant, Smail, Swinbank, Valiante, Verma, Vieira, Wardlow,
  \& van~der Werf}]{Fu:2012iu}
Fu, H., Jullo, E., Cooray, A., {et~al.} 2012, ApJ, 753, 134

\bibitem[{Fu {et~al.}(2013)Fu, Cooray, Feruglio, Ivison, Riechers, Gurwell,
  Bussmann, Harris, Altieri, Aussel, Baker, Bock, Boylan-Kolchin, Bridge,
  Calanog, Casey, Cava, Chapman, Clements, Conley, Cox, Farrah, Frayer,
  Hopwood, Jia, Magdis, Marsden, Mart{\'\i}nez-Navajas, Negrello, Neri, Oliver,
  Omont, Page, P{\'e}rez-Fournon, Schulz, Scott, Smith, Vaccari, Valtchanov,
  Vieira, Viero, Wang, Wardlow, \& Zemcov}]{Fu:2013hz}
Fu, H., Cooray, A., Feruglio, C., {et~al.} 2013, Nature, 498, 338

\bibitem[{Genzel {et~al.}(2012)Genzel, Tacconi, Combes, Bolatto, Neri,
  Sternberg, Cooper, Bouch{\'e}, Bournaud, Burkert, Comerford, Cox, Davis,
  F{\"o}rster~Schreiber, Garcia-Burillo, Gracia-Carpio, Lutz, Naab, Newman,
  Saintonge, Shapiro, Shapley, \& Weiner}]{Genzel:2012fw}
Genzel, R., Tacconi, L.~J., Combes, F., {et~al.} 2012, ApJ, 746, 69

\bibitem[{Genzel {et~al.}(2017)Genzel, Schreiber, {\"U}bler, Lang, Naab,
  Bender, Tacconi, Wisnioski, Wuyts, Alexander, Beifiori, Belli, Brammer,
  Burkert, Carollo, Chan, Davies, Fossati, Galametz, Genel, Gerhard, Lutz,
  Mendel, Momcheva, Nelson, Renzini, Saglia, Sternberg, Tacchella, Tadaki, \&
  Wilman}]{Genzel:2017fa}
Genzel, R., Schreiber, N. M.~F., {\"U}bler, H., {et~al.} 2017, Nature, 543, 397

\bibitem[{Goodman \& Weare(2010)}]{Goodman:2010et}
Goodman, J., \& Weare, J. 2010, Communications in Applied Mathematics and
  Computational Science, Vol. 5, No. 1, p. 65-80, 2010, 5, 65

\bibitem[{Hodge {et~al.}(2012)Hodge, Carilli, Walter, de~Blok, Riechers, Daddi,
  \& Lentati}]{Hodge:2012jt}
Hodge, J.~A., Carilli, C., Walter, F., {et~al.} 2012, ApJ, 760, 11

\bibitem[{Hughes {et~al.}(1998)Hughes, Serjeant, Dunlop, Rowan-Robinson, Blain,
  Mann, Ivison, Peacock, Efstathiou, Gear, Oliver, Lawrence, Longair,
  Goldschmidt, \& Jenness}]{Hughes:1998gn}
Hughes, D.~H., Serjeant, S., Dunlop, J., {et~al.} 1998, Nature, 394, 241

\bibitem[{Ivison {et~al.}(2011)Ivison, Papadopoulos, Smail, Greve, Thomson,
  Xilouris, \& Chapman}]{Ivison:2011it}
Ivison, R.~J., Papadopoulos, P.~P., Smail, I., {et~al.} 2011, MNRAS, 412, 1913

\bibitem[{Ivison {et~al.}(1998)Ivison, Smail, Le~Borgne, Blain, Kneib,
  Bezecourt, Kerr, \& Davies}]{Ivison:1998iw}
Ivison, R.~J., Smail, I., Le~Borgne, J.~F., {et~al.} 1998, MNRAS, 298, 583

\bibitem[{Ivison {et~al.}(2010)Ivison, Smail, Papadopoulos, Wold, Richard,
  Swinbank, Kneib, \& Owen}]{Ivison:2010kz}
Ivison, R.~J., Smail, I., Papadopoulos, P.~P., {et~al.} 2010, MNRAS, 404, 198

\bibitem[{Ivison {et~al.}(2013)Ivison, Swinbank, Smail, Harris, Bussmann,
  Cooray, Cox, Fu, Kov{\'a}cs, Krips, Narayanan, Negrello, Neri,
  Pe{\~n}arrubia, Richard, Riechers, Rowlands, Staguhn, Targett, Amber, Baker,
  Bourne, Bertoldi, Bremer, Calanog, Clements, Dannerbauer, Dariush, de~Zotti,
  Dunne, Eales, Farrah, Fleuren, Franceschini, Geach, George, Helly, Hopwood,
  Ibar, Jarvis, Maddox, Omont, Scott, Serjeant, Smith, Thompson, Valiante,
  Valtchanov, Vieira, \& van~der Werf}]{Ivison:2013hj}
Ivison, R.~J., Swinbank, A.~M., Smail, I., {et~al.} 2013, ApJ, 772, 137

\bibitem[{J{\'o}zsa {et~al.}(2007)J{\'o}zsa, Kenn, Klein, \&
  Oosterloo}]{Jozsa:2007ey}
J{\'o}zsa, G. I.~G., Kenn, F., Klein, U., \& Oosterloo, T.~A. 2007, A{\&}A,
  468, 731

\bibitem[{Komatsu {et~al.}(2011)Komatsu, Smith, Dunkley, Bennett, Gold,
  Hinshaw, Jarosik, Larson, Nolta, Page, Spergel, Halpern, Hill, Kogut, Limon,
  Meyer, Odegard, Tucker, Weiland, Wollack, \& Wright}]{Komatsu:2011in}
Komatsu, E., Smith, K.~M., Dunkley, J., {et~al.} 2011, ApJS, 192, 18

\bibitem[{Lang {et~al.}(2017)Lang, F{\"o}rster~Schreiber, Genzel, Wuyts,
  Wisnioski, Beifiori, Belli, Bender, Brammer, Burkert, Chan, Davies, Fossati,
  Galametz, Kulkarni, Lutz, Mendel, Momcheva, Naab, Nelson, Saglia, Seitz,
  Tacchella, Tacconi, Tadaki, {\"U}bler, van Dokkum, \& Wilman}]{Lang:2017bv}
Lang, P., F{\"o}rster~Schreiber, N.~M., Genzel, R., {et~al.} 2017, ApJ, 840, 92

\bibitem[{Levy {et~al.}(2018)Levy, Bolatto, Teuben, Sanchez,
  Barrera-Ballesteros, Blitz, Colombo, Garcia-Benito, Herrera-Camus, Husemann,
  Kalinova, Lan, Leung, Mast, Utomo, van~de Ven, Vogel, \& Wong}]{Levy:2018hj}
Levy, R.~C., Bolatto, A.~D., Teuben, P., {et~al.} 2018, ApJ, 860, 92

\bibitem[{Liu {et~al.}(2017)Liu, Weiss, Perez-Beaupuits, G{\"u}sten, Liu, Gao,
  Menten, Werf, Israel, Harris, Martin-Pintado, Requena-Torres, \&
  Stutzki}]{Liu:2017ce}
Liu, L., Weiss, A., Perez-Beaupuits, J.~P., {et~al.} 2017, ApJ, 846, 5

\bibitem[{Magnelli {et~al.}(2012)Magnelli, Saintonge, Lutz, Tacconi, Berta,
  Bournaud, Charmandaris, Dannerbauer, Elbaz, F{\"o}rster~Schreiber,
  Gracia-Carpio, Ivison, Maiolino, Nordon, Popesso, Rodighiero, Santini, \&
  Wuyts}]{Magnelli:2012br}
Magnelli, B., Saintonge, A., Lutz, D., {et~al.} 2012, A{\&}A, 548, 22

\bibitem[{McMullin {et~al.}(2007)McMullin, Waters, Schiebel, Young, \&
  Golap}]{McMullin:2007ta}
McMullin, J.~P., Waters, B., Schiebel, D., Young, W., \& Golap, K. 2007, adass
  XVIII, 376, 127

\bibitem[{Mineo {et~al.}(2012)Mineo, Gilfanov, \& Sunyaev}]{Mineo:2012kz}
Mineo, S., Gilfanov, M., \& Sunyaev, R. 2012, MNRAS, 419, 2095

\bibitem[{Narayanan {et~al.}(2012)Narayanan, Krumholz, Ostriker, \&
  Hernquist}]{Narayanan:2012fa}
Narayanan, D., Krumholz, M.~R., Ostriker, E.~C., \& Hernquist, L. 2012, MNRAS,
  421, 3127

\bibitem[{Nayyeri {et~al.}(2016)Nayyeri, Keele, Cooray, Riechers, Ivison,
  Harris, Frayer, Baker, Chapman, Eales, Farrah, Fu, Marchetti, Marques-Chaves,
  Martinez-Navajas, Oliver, Omont, Perez-Fournon, Scott, Vaccari, Vieira,
  Viero, Wang, \& Wardlow}]{Nayyeri:2016iy}
Nayyeri, H., Keele, M., Cooray, A., {et~al.} 2016, ApJ, 823, 17

\bibitem[{Negrello {et~al.}(2010)Negrello, Hopwood, de~Zotti, Cooray, Verma,
  Bock, Frayer, Gurwell, Omont, Neri, Dannerbauer, Leeuw, Barton, Cooke, Kim,
  da~Cunha, Rodighiero, Cox, Bonfield, Jarvis, Serjeant, Ivison, Dye, Aretxaga,
  Hughes, Ibar, Bertoldi, Valtchanov, Eales, Dunne, Driver, Auld, Buttiglione,
  Cava, Grady, Clements, Dariush, Fritz, Hill, Hornbeck, Kelvin, Lagache,
  Lopez-Caniego, Gonzalez-Nuevo, Maddox, Pascale, Pohlen, Rigby, Robotham,
  Simpson, Smith, Temi, Thompson, Woodgate, York, Aguirre, Beelen, Blain,
  Baker, Birkinshaw, Blundell, Bradford, Burgarella, Danese, Dunlop, Fleuren,
  Glenn, Harris, Kamenetzky, Lupu, Maddalena, Madore, Maloney, Matsuhara,
  Michalowski, Murphy, Naylor, Nguyen, Popescu, Rawlings, Rigopoulou, Scott,
  Scott, Seibert, Smail, Tuffs, Vieira, van~der Werf, \&
  Zmuidzinas}]{Negrello:2010fx}
Negrello, M., Hopwood, R., de~Zotti, G., {et~al.} 2010, Science, 330, 800

\bibitem[{Negrello {et~al.}(2017)Negrello, Amber, Amvrosiadis, Cai, Lapi,
  Gonzalez-Nuevo, de~Zotti, Furlanetto, Maddox, Allen, Bakx, Bussmann, Cooray,
  Covone, Danese, Dannerbauer, Fu, Greenslade, Gurwell, Hopwood, Koopmans,
  Napolitano, Nayyeri, Omont, Petrillo, Riechers, Serjeant, Tortora, Valiante,
  Verdoes~Kleijn, Vernardos, Wardlow, Baes, Baker, Bourne, Clements, Crawford,
  Dye, Dunne, Eales, Ivison, Marchetti, Michalowski, Smith, Vaccari, \& van~der
  Werf}]{Negrello:2017jp}
Negrello, M., Amber, S., Amvrosiadis, A., {et~al.} 2017, MNRAS, 465, 3558

\bibitem[{Noordermeer {et~al.}(2007)Noordermeer, van~der Hulst, Sancisi,
  Swaters, \& van Albada}]{Noordermeer:2007gm}
Noordermeer, E., van~der Hulst, J.~M., Sancisi, R., Swaters, R.~S., \& van
  Albada, T.~S. 2007, MNRAS, 376, 1513

\bibitem[{Ojha {et~al.}(2001)Ojha, Stark, Hsieh, Lane, Chamberlin, Bania,
  Bolatto, Jackson, \& Wright}]{Ojha:2001gs}
Ojha, R., Stark, A.~A., Hsieh, H.~H., {et~al.} 2001, ApJ, 548, 253

\bibitem[{Omont {et~al.}(2013)Omont, Yang, Cox, Neri, Beelen, Bussmann,
  Gavazzi, van~der Werf, Riechers, Downes, Krips, Dye, Ivison, Vieira, Weiss,
  Aguirre, Baes, Baker, Bertoldi, Cooray, Dannerbauer, de~Zotti, Eales, Fu,
  Gao, Gu{\'e}lin, Harris, Jarvis, Lehnert, Leeuw, Lupu, Menten,
  Micha{\l}owski, Negrello, Serjeant, Temi, Auld, Dariush, Dunne, Fritz,
  Hopwood, Hoyos, Ibar, Maddox, Smith, Valiante, Bock, Bradford, Glenn, \&
  Scott}]{Omont:2013kc}
Omont, A., Yang, C., Cox, P., {et~al.} 2013, A{\&}A, 551, A115

\bibitem[{Papadopoulos {et~al.}(2004)Papadopoulos, Thi, \&
  Viti}]{Papadopoulos:2004jl}
Papadopoulos, P.~P., Thi, W.-F., \& Viti, S. 2004, MNRAS, 351, 147

\bibitem[{Papadopoulos {et~al.}(2012)Papadopoulos, van~der Werf, Xilouris,
  Isaak, \& Gao}]{Papadopoulos:2012hj}
Papadopoulos, P.~P., van~der Werf, P., Xilouris, E., Isaak, K.~G., \& Gao, Y.
  2012, ApJ, 751, 10

\bibitem[{Perley \& Butler(2013)}]{Perley:2013fa}
Perley, R.~A., \& Butler, B.~J. 2013, ApJS, 204, 19

\bibitem[{Pilbratt {et~al.}(2010)Pilbratt, Riedinger, Passvogel, Crone, Doyle,
  Gageur, Heras, Jewell, Metcalfe, Ott, \& Schmidt}]{Pilbratt:2010en}
Pilbratt, G.~L., Riedinger, J.~R., Passvogel, T., {et~al.} 2010, A{\&}A, 518,
  L1

\bibitem[{Rangwala {et~al.}(2011)Rangwala, Maloney, Glenn, Wilson, Rykala,
  Isaak, Baes, Bendo, Boselli, Bradford, Clements, Cooray, Fulton, Imhof,
  Kamenetzky, Madden, Mentuch, Sacchi, Sauvage, Schirm, Smith, Spinoglio, \&
  Wolfire}]{Rangwala:2011dd}
Rangwala, N., Maloney, P.~R., Glenn, J., {et~al.} 2011, ApJ, 743, 94

\bibitem[{Riechers {et~al.}(2010)Riechers, Capak, Carilli, Cox, Neri, Scoville,
  Schinnerer, Bertoldi, \& Yan}]{Riechers:2010kf}
Riechers, D.~A., Capak, P.~L., Carilli, C.~L., {et~al.} 2010, ApJ, 720, L131

\bibitem[{Rubin \& Ford(1970)}]{Rubin:1970gu}
Rubin, V.~C., \& Ford, W. K.~J. 1970, ApJ, 159, 379

\bibitem[{Sch{\"o}ier {et~al.}(2005)Sch{\"o}ier, van~der Tak, van Dishoeck, \&
  Black}]{Schoier:2005wx}
Sch{\"o}ier, F.~L., van~der Tak, F. F.~S., van Dishoeck, E.~F., \& Black, J.~H.
  2005, A{\&}A, 432, 369

\bibitem[{Scoville(2004)}]{Scoville:2004uf}
Scoville, N. 2004, adass XVIII, 320, 253

\bibitem[{Smail {et~al.}(1997)Smail, Ivison, \& Blain}]{Smail:1997kh}
Smail, I., Ivison, R.~J., \& Blain, A.~W. 1997, ApJ, 490, L5

\bibitem[{Swaters {et~al.}(2009)Swaters, Sancisi, van Albada, \& van~der
  Hulst}]{Swaters:2009ku}
Swaters, R.~A., Sancisi, R., van Albada, T.~S., \& van~der Hulst, J.~M. 2009,
  A{\&}A, 493, 871

\bibitem[{Tacconi {et~al.}(2008)Tacconi, Genzel, Smail, Neri, Chapman, Ivison,
  Blain, Cox, Omont, Bertoldi, Greve, F{\"o}rster~Schreiber, Genel, Lutz,
  Swinbank, Shapley, Erb, Cimatti, Daddi, \& Baker}]{Tacconi:2008fe}
Tacconi, L.~J., Genzel, R., Smail, I., {et~al.} 2008, ApJ, 680, 246

\bibitem[{Thompson {et~al.}(2005)Thompson, Quataert, \&
  Murray}]{Thompson:2005dq}
Thompson, T.~A., Quataert, E., \& Murray, N. 2005, ApJ, 630, 167

\bibitem[{van~der Kruit \& Allen(1978)}]{vanderKruit:1978cn}
van~der Kruit, P., \& Allen, R.~J. 1978, ARA{\&}A, 16, 103

\bibitem[{van~der Tak {et~al.}(2007)van~der Tak, Black, Sch{\"o}ier, Jansen, \&
  van Dishoeck}]{vanderTak:2007be}
van~der Tak, F. F.~S., Black, J.~H., Sch{\"o}ier, F.~L., Jansen, D.~J., \& van
  Dishoeck, E.~F. 2007, A{\&}A, 468, 627

\bibitem[{van~der Wel {et~al.}(2014)van~der Wel, Franx, van Dokkum, Skelton,
  Momcheva, Whitaker, Brammer, Bell, Rix, Wuyts, Ferguson, Holden, Barro,
  Koekemoer, Chang, McGrath, H{\"a}u{\ss}ler, Dekel, Behroozi, Fumagalli, Leja,
  Lundgren, Maseda, Nelson, Wake, Patel, Labb{\'e}, Faber, Grogin, \&
  Kocevski}]{vanderWel:2014hi}
van~der Wel, A., Franx, M., van Dokkum, P.~G., {et~al.} 2014, ApJ, 788, 28

\bibitem[{Wardlow {et~al.}(2013)Wardlow, Cooray, De~Bernardis, Amblard,
  Arumugam, Aussel, Baker, B{\'e}thermin, Blundell, Bock, Boselli, Bridge,
  Buat, Burgarella, Bussmann, Cabrera-Lavers, Calanog, Carpenter, Casey,
  Castro-Rodriguez, Cava, Chanial, Chapin, Chapman, Clements, Conley, Cox,
  Dowell, Dye, Eales, Farrah, Ferrero, Franceschini, Frayer, Frazer, Fu,
  Gavazzi, Glenn, Gonz{\'a}lez~Solares, Griffin, Gurwell, Harris,
  Hatziminaoglou, Hopwood, Hyde, Ibar, Ivison, Kim, Lagache, Levenson,
  Marchetti, Marsden, Mart{\'\i}nez-Navajas, Negrello, Neri, Nguyen,
  O'Halloran, Oliver, Omont, Page, Panuzzo, Papageorgiou, Pearson,
  Perez-Fournon, Pohlen, Riechers, Rigopoulou, Roseboom, Rowan-Robinson,
  Schulz, Scott, Scoville, Seymour, Shupe, Smith, Streblyanska, Strom,
  Symeonidis, Trichas, Vaccari, Vieira, Viero, Wang, Xu, Yan, \&
  Zemcov}]{Wardlow:2013ie}
Wardlow, J.~L., Cooray, A., De~Bernardis, F., {et~al.} 2013, ApJ, 762, 59

\bibitem[{Weiss {et~al.}(2005)Weiss, Walter, \& Scoville}]{Weiss:2005fk}
Weiss, A., Walter, F., \& Scoville, N.~Z. 2005, A{\&}A, 438, 533

\bibitem[{Wisnioski {et~al.}(2015)Wisnioski, F{\"o}rster~Schreiber, Wuyts,
  Wuyts, Bandara, Wilman, Genzel, Bender, Davies, Fossati, Lang, Mendel,
  Beifiori, Brammer, Chan, Fabricius, Fudamoto, Kulkarni, Kurk, Lutz, Nelson,
  Momcheva, Rosario, Saglia, Seitz, Tacconi, \& van Dokkum}]{Wisnioski:2015gk}
Wisnioski, E., F{\"o}rster~Schreiber, N.~M., Wuyts, S., {et~al.} 2015, ApJ,
  799, 209

\bibitem[{Yang {et~al.}(2013)Yang, Gao, Omont, Liu, Isaak, Downes, van~der
  Werf, \& Lu}]{Yang:2013dj}
Yang, C., Gao, Y., Omont, A., {et~al.} 2013, ApJ, 771, L24

\bibitem[{Yang {et~al.}(2016)Yang, Omont, Beelen, Gonzalez-Alfonso, Neri, Gao,
  van~der Werf, Weiss, Gavazzi, Falstad, Baker, Bussmann, Cooray, Cox,
  Dannerbauer, Dye, Gu{\'e}lin, Ivison, Krips, Lehnert, Micha{\l}owski,
  Riechers, Spaans, \& Valiante}]{Yang:2016kr}
Yang, C., Omont, A., Beelen, A., {et~al.} 2016, A{\&}A, 595, A80

\bibitem[{Zhang {et~al.}(2018)Zhang, Romano, Ivison, Papadopoulos, \&
  Matteucci}]{Zhang:2018ee}
Zhang, Z.-Y., Romano, D., Ivison, R.~J., Papadopoulos, P.~P., \& Matteucci, F.
  2018, Nature, 558, 260

\end{thebibliography}

\end{document}